\documentstyle [preprint,eqsecnum,aps,epsfig]{revtex}

\begin{document}

\count255=\time\divide\count255 by 60 \xdef\hourmin{\number\count255}
  \multiply\count255 by-60\advance\count255 by\time
 \xdef\hourmin{\hourmin:\ifnum\count255<10 0\fi\the\count255}

\draft\preprint{WM-01-110}

\title{Bounding Noncommutative QCD}

\author{Carl E. Carlson$^a$\footnote{carlson@physics.wm.edu},
Christopher D. Carone$^a$\footnote{carone@physics.wm.edu}, and
Richard F. Lebed$^b$\footnote{Richard.Lebed@asu.edu}}

\vskip 0.1in

\address{$^a$Nuclear and Particle Theory Group, Department of
Physics, College of William and Mary, Williamsburg, VA 23187-8795}

\vskip 0.2in

\address{$^b$Department of Physics and Astronomy, Arizona State University,
Tempe, AZ 85287-1504}

\vskip .1in
\date{July, 2001}
\vskip .1in

\maketitle
\thispagestyle{empty}

\begin{abstract}
Jur\v{c}o, M\"{o}ller, Schraml, Schupp, and Wess have shown how to 
construct noncommutative SU(N) gauge theories from a consistency relation.  
Within this framework, we present the Feynman rules for noncommutative QCD 
and compute explicitly the most dangerous Lorentz-violating operator 
generated through radiative corrections.  We find that interesting effects 
appear at the one-loop level, in contrast to conventional noncommutative 
U(N) gauge theories, leading to a stringent bound. Our results are 
consistent with others appearing recently in the literature that suggest 
collider limits are not competitive with low-energy tests of Lorentz 
violation for bounding the scale of spacetime noncommutativity.
\end{abstract}


\newpage
\setcounter{page}{1}

\section{Introduction}\label{sec:intro}

Recent interest in the possibility of extra spatial dimensions
with large radii of compactification~\cite{led} has led a number of 
groups to consider other modifications of spacetime structure that 
may have observable consequences. One such possibility is that the usual
four spacetime dimensions become noncommutative at some scale
$\Lambda_{{\rm NC}}$  higher than currently accessible energies, {\it
viz.,}
\begin{equation}
[\hat{x}^\mu,\hat{x}^\nu]= i \theta^{\mu\nu} \,\,\, ,
\label{eq:comrel}
\end{equation}
where a hat indicates a noncommuting coordinate, and where
$\theta$ is a real, constant matrix with elements of order
$(\Lambda_{{\rm NC}})^{-2}$~\cite{Hew,Hinch,Anis,Moc,Carr,Chaichian:2001si}.
Such noncommutative geometries have been shown to arise in the low-energy 
limit of some string theories~\cite{Seiberg}, and lead to quantum field 
theories that are interesting in their own 
right~\cite{Haya,Madore,Armoni,Jurco,Chaichian:2001py}. Experimental 
signals of noncommutativity have been discussed from the point of view of 
collider physics~\cite{Hew,Hinch} as well as low-energy non-accelerator 
experiments~\cite{Anis,Moc,Carr,Chaichian:2001si}. Two widely disparate 
sets of bounds on $\Lambda_{{\rm NC}}$ can be found in the literature: 
bounds of order a TeV from colliders~\cite{Hew,Hinch}, and bounds of 
order $10^{11}$~GeV \cite{Anis} or higher~\cite{Moc} from low-energy tests of 
Lorentz violation.  Bounds of the latter type will be of interest to us in 
this letter.

The Lorentz violation in noncommutative field theories originates
from the constants $\theta^{\mu\nu}$ appearing in
Eq.~(\ref{eq:comrel}). Noting that $\theta^{\mu\nu}$ is
antisymmetric, one can regard $\theta^{0i}$ and $\theta^{ij}$ as
constant three-vectors indicating preferred directions in a given
Lorentz frame.  How $\theta$ appears in the low-energy effective
theory determines the ways in which Lorentz violation may be
manifested in experiment.  As pointed out by Mocioiu {\em et al.}~\cite{Moc},
the existence of an operator of the form
\begin{equation}
\theta^{\mu \nu} \overline{q} \sigma_{\mu \nu} q, \label{eq:theop}
\end{equation}
where $q$ is a quark field, leads to a shift in nuclear magnetic
moments and an observable sidereal variation in the magnitude of
hyperfine splitting in atoms~\cite{Berglund95}.  From here, they
obtain the bound $\Lambda_{{\rm NC}} > 5 \times 10^{14}$~GeV.  
Anisimov, Banks, Dine, and Graesser~\cite{Anis} have shown that an 
operator of this form is generated via quantum corrections in a number 
of toy models.  Both Refs.~\cite{Anis} and~\cite{Moc} are speculative in 
the sense that the relevant calculation was not done in a well-defined 
noncommutative generalization of QCD. This stems from long-standing 
difficulties in formulating noncommutative SU(N) gauge 
theories~\cite{Madore,Armoni}. (Similar difficulties arise in constructing 
noncommutative U(1) theories with fractionally-charged matter 
fields~\cite{Haya}.)  Recently, Jur\v{c}o {\em et al.}~\cite{Jurco} have 
demonstrated how properly to construct a noncommutative SU(N)  gauge theory 
by implementing consistency conditions order by order in the 
noncommutativity parameter $\theta^{\mu\nu}$. We review this formulation 
below.  The purpose of our work is two-fold: we first present the Feynman 
rules derived from this consistent formulation of 
noncommutative QCD, which differ from those that have appeared 
previously in the literature~\cite{Hinch}.  Using these results, we show 
that an operator of the same form as Eq.~(\ref{eq:theop}) is indeed 
generated. From here we can immediately place a bound on the scale of 
spacetime noncommutativity.  

In the conventional formulation of noncommutative gauge theory, one works 
with an equivalent theory of quantum fields that are functions of commuting 
spacetime coordinates by promoting ordinary multiplication to a Moyal star 
product,
\begin{equation}
f(\hat{x})g(\hat{x}) \rightarrow f(x)\star g(x)
\end{equation}
where
\begin{equation}\label{eq:stardef}
f(x)\star g(x)=e^{\frac{i}{2} \frac{\partial}{\partial x^\mu} \theta^{\mu\nu}
\frac{\partial}{\partial y^\nu}} f(x) g(y)|_{y\rightarrow x} \,\,\, .
\end{equation}
For a non-Abelian gauge group, with generators $T^a$ and a Lie algebra-valued
gauge parameter $\alpha \equiv \alpha^a T^a$, one expects the gauge 
field $\hat{A}^\mu \equiv A^{\mu a}(\hat{x}) T^a$ to transform 
infinitesimally as
\begin{equation}\label{eq:deltaa}
\delta \hat{A}^\mu = \partial^\mu \alpha + i [\alpha, \hat{A}^\mu] \,\,\, .
\end{equation}
In particular, the commutator can be expanded in terms of the
group generators,
\begin{equation}\label{eq:prob}
[\alpha,\hat{A}] = \frac{1}{2}(\alpha_r \hat{A}_s +\hat{A}_s \alpha_r) 
[T^r,T^s]
+\frac{1}{2} (\alpha_r \hat{A}_s - \hat{A}_s \alpha_r) \{T^r,T^s\}  \,\,\,  .
\end{equation}
The problem in formulating an SU(N) gauge theory is now manifest:
while the first term lives within an SU(N) representation, the
second term does not. For example, in the case of SU(2), the second
term is proportional to the identity, and tracelessness of
the gauge generators is not maintained.

This problem has led others to focus on U(N) groups to
approximate the phenomenology that might be relevant in a
noncommutative SU(N) theory~\cite{Hinch,Anis}.  We will instead 
focus on a consistent formulation of noncommutative SU(N), and study the
associated phenomenology directly.

\section{Noncommutative SU(N)}

The stumbling block presented by Eq.~(\ref{eq:prob}) is that it 
implies $\delta A$ lives in the enveloping algebra of the
gauge group of interest.  Jur\v{c}o {\em et al.} show that it is
nonetheless possible to define gauge transformations in this
larger algebra that depend only on the gauge parameter and fields
of the desired theory.    For concreteness, consider
a noncommutative SU(N) gauge theory in which the fields transform
infinitesimally as
\begin{equation} \label{eq:trans}
\delta_\alpha \psi = i \Lambda_\alpha \star \psi
\,\,\,\,\,\,\,\,\,\, , \,\,\,\,\,\,\,\,\,\, \delta_\alpha A_\mu =
\partial_\mu \Lambda_\alpha + i [\Lambda_\alpha\stackrel{\star}{,}A_\mu] 
\,\,\, .
\end{equation}
Here, $\Lambda_\alpha$ is a U(N) matrix function that we wish to associate
with an element of SU(N) corresponding to the gauge parameter
$\alpha$.  The appropriate consistency condition is
\begin{equation}
(\delta_\alpha\delta_\beta-\delta_\beta\delta_\alpha) \psi(x)
=\delta_{\alpha \times \beta} \psi(x) \,\,\, ,
\end{equation}
where $\alpha\times\beta$ represents $\alpha_a \beta_b f^{abc} T^c$, with
$f^{abc}$ and $T^a$ the structure constants and generators of
SU(N), respectively.  With some algebra, one may show that this
constraint is satisfied if
\begin{equation}
\Lambda_\alpha[A^0] = \alpha + \frac{1}{4} \theta^{\mu\nu}
\{ \partial_\mu \alpha\mbox{ , }A_\nu^0\} + {O}(\theta^2) \,\,\, ,
\end{equation}
where $A_\nu^0 \equiv A_\nu^{0a} T^a$ is the usual SU(N) gauge field.  The
formulation of Jur\v{c}o {\em et al.} requires the gauge
parameter $\Lambda_\alpha$ to be a nontrivial function of the gauge
field $A^0$.  The same holds true for the fields $\psi$ and $A$:
the requirement that the matter and gauge fields transform as in
Eq.~(\ref{eq:trans}) implies that
\begin{equation}
A^\mu = A^{0\mu}-\frac{1}{4}\theta_{\rho\nu} \{ A^{0\rho}\!\!
\mbox{ , }\,
\partial^\nu A^{0\mu}+F^{0\nu\mu}\}
\end{equation}
and
\begin{equation}\label{eq:fields}
\psi = \psi^0 -\frac{1}{2}\theta^{\mu\nu}A_\mu^0\partial_\nu\psi^0
+\frac{i}{4}\theta^{\mu\nu} A^0_\mu A^0_\nu \psi^0  \,\,\, ,
\end{equation}
to linear order in $\theta$.  While $A^0$ and $\psi^0$ have
the usual transformation properties of fields in an SU(N) gauge
theory, the Lagrangian expressed in terms of these fields
is different.  Starting with the action
\begin{equation}\label{eq:firstaction}
S=\int d^4 x \left[ \bar \psi \star (i \! \not \! \! {\cal D} - m )\psi -
\frac{1}{2g^2} \, {\rm Tr} \, F_{\mu \nu} \! \star F^{\mu \nu} \right]
\,\,\, ,
\end{equation}
in which
\begin{equation}
{\cal D}_\mu \psi  \equiv  \partial_\mu \psi - i A_\mu \star
\psi \,\,\, \mbox{,} \,\,\,  F_{\mu \nu}  \equiv  \partial_\mu
A_\nu - \partial_\nu A_\mu - i [ A_\mu \stackrel{\star}{,} A_\nu ] 
\,\,\, ,
\end{equation}
one may expand in terms of $\psi^0$, $A^0_\mu$, and $\theta$:
\[
S = \int d^4 x \bigg[ \bar \psi^0 (i \! \not \! \! {\cal D} - m
)\psi^0 - \frac 1 4 \, \theta^{\mu \nu} \bar \psi^0 F^0_{\mu \nu}
(i \not \! \! {\cal D} - m )\psi^0 - \frac 1 2 \, \theta^{\mu \nu}
\bar \psi^0 \gamma^\rho F^0_{\rho \mu} \, i {\cal D}_\nu \psi^0
 \]
\begin{equation}\label{eq:action}
 -\frac{1}{2g^2} \, {\rm Tr} \, F^0_{\mu \nu} F^{0 \mu \nu}
+ \frac{1}{4g^2} \, \theta^{\mu \nu} \, {\rm Tr} \, F^0_{\mu \nu}
F^0_{\rho \sigma} F^{0 \rho \sigma} -\frac{1}{g^2} \, \theta^{\mu
\nu} \, {\rm Tr} \, F^0_{\mu \rho} F^0_{\nu \sigma} F^{0 \rho
\sigma} \bigg] .
\end{equation}
Here we have corrected trivial typographical errors that appear in
Ref.~\cite{Jurco} and adopted  Bjorken and Drell conventions~\cite{BD} for 
the gamma matrices.  Note that derivatives and field strengths now have 
their ordinary meaning,
\begin{equation}
{\cal D}_\mu \psi^0  \equiv  \partial_\mu \psi^0 - i A^0_\mu
\psi^0 \,\,\, \mbox{,} \,\,\,  F_{\mu \nu}^0  \equiv  \partial_\mu
A_\nu^0 -
\partial_\nu A_\mu^0 - i [ A_\mu^0, A_\nu^0 ] \,\,\, .
\end{equation}
Equation~(\ref{eq:action}) is, to order $\theta$, a Seiberg-Witten 
map~\cite{Seiberg} for noncommutative SU(N), {\em i.e.}, a physically 
equivalent theory written in terms of fields that are functions of 
commuting spacetime coordinates.

\begin{figure}[ht]
\centerline{\epsfbox{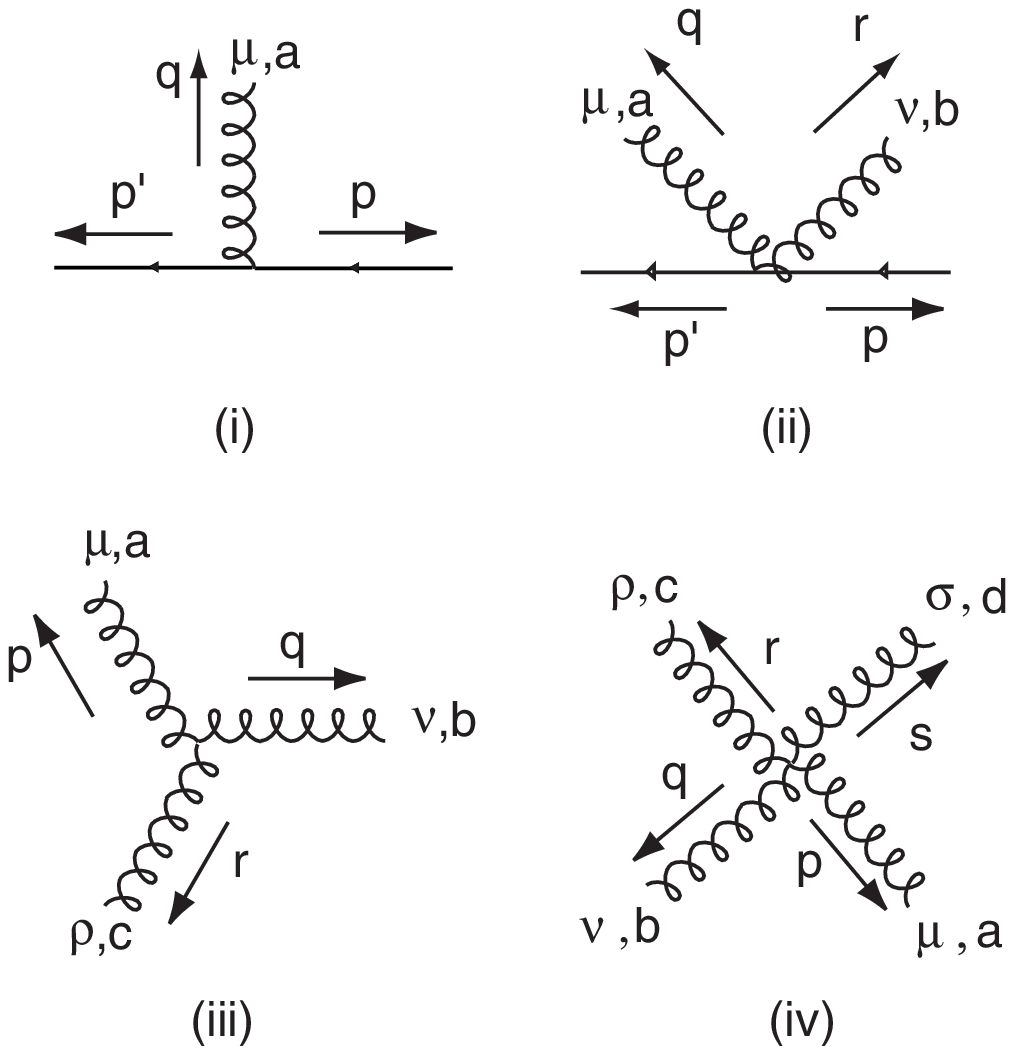}}
\caption{Feynman vertices up to $O(g^2 \theta^1)$.}
\label{rules}
\end{figure}

Feynman rules may be extracted from Eq.~(\ref{eq:action}) in the
usual way.  The complete ${O}(\theta^1)$ Lagrangian is seen to
contain a variety of interaction vertices, namely, those with two
fermions and 1, 2, and 3 gauge bosons, and pure gauge vertices with 3,
4, 5, and 6 bosons.  Here we present the Feynman rules for vertices
up to $O(g^2 \theta^1)$; this includes the new quark-quark-gluon 
vertex that will be relevant to the calculation presented
in the next section.  Our conventions are as follows:  The structure
constants for SU(N) are defined by
\begin{equation}
[ T^a, T^b ] = i f^{abc} \, T^c \,\,\,\,\, \mbox{ and } \,\,\,\,\,
\{ T^a, T^b \} = d^{abc} T^c + \frac{1}{N} \delta^{ab} {\bf \openone} \,\,\, .
\end{equation}
In addition, we define the following abbreviations for tensor
contractions: $\theta^\mu \cdot  p \equiv \theta^{\mu \nu} p_\nu$
and $p \cdot \theta \cdot q \equiv \theta^{\mu \nu} p_\mu q_\nu$,
for any four-vectors $p, q$. Finally, it is convenient to introduce
the totally antisymmetric three-index symbol
\begin{equation}
\theta^{\mu \nu \rho} \equiv \theta^{\mu \nu} \gamma^\rho +
\theta^{\nu \rho} \gamma^\mu + \theta^{\rho \mu} \gamma^\nu .
\end{equation}
The Feynman rules for the ${O}(\theta^1)$ contributions to
the vertices shown in Fig.~\ref{rules} are

$qqg$ vertex (i):
\begin{equation}
\frac{g}{2} T^a \left[\theta^\mu\cdot p \, (\not\! p'- \,m) -
\theta^\mu\cdot p' (\not\! p+ \,m)-p'\cdot\theta\cdot p \,
\gamma^\mu \right]
\end{equation}

$qqgg$ vertex (ii):
\begin{equation}
\frac{g^2}{2}\left\{ T^a
T^b [m\,\theta^{\mu\nu}+\theta^{\mu\nu\rho} (p+q)_\rho]-T^b T^a [m\,
\theta^{\mu\nu}+\theta^{\mu\nu\rho} (p+r)_\rho]\right\}
\end{equation}

$ggg$ vertex (iii):
\begin{eqnarray}
-\frac{1}{2} g d_{abc} \left\{ r\cdot\theta\cdot q \left[
(q-r)^\mu g^{\nu\rho}+(p-q)^\rho g^{\mu\nu}+(r-p)^\nu
g^{\mu\rho} \right] \right.
\nonumber \\
+(q^2 g^{\rho\nu}-q^\rho q^\nu)\theta^\mu\cdot r +(r^2
g^{\rho\nu}-r^\rho r^\nu)\theta^\mu\cdot q+(q^2 g^{\mu\nu}-q^\mu
q^\nu)\theta^\rho\cdot p
\nonumber \\
+(p^2 g^{\mu\nu}-p^\mu p^\nu)\theta^\rho\cdot q+(r^2
g^{\mu\rho}-r^\mu r^\rho)\theta^\nu\cdot p + (p^2
g^{\mu\rho}-p^\mu p^\rho)\theta^\nu\cdot r
\nonumber \\
+\left. ( q\cdot p \, r^\nu-r\cdot q \, p^\nu)\theta^{\mu \rho}
+(r\cdot q \, p ^\rho - p\cdot r \, q^\rho)\theta^{\nu\mu}
+(p\cdot r \, q^\mu-q\cdot p \, r^\mu)\theta^{\rho \nu} \right\}
\end{eqnarray}

$gggg$ vertex (iv):
\begin{eqnarray}
 -i \frac{g^2}{2} f^{abe}d^{cde}
\left\{ \theta^{\mu\nu}(g^{\rho\sigma} r\cdot s -r^\sigma s^\rho) +
\theta^{\rho\sigma}(r^\nu s^\mu- r^\mu s^\nu)-\theta^{\mu\rho}
(g^{\nu\sigma}r\cdot s-r^\sigma s^\nu)\right.
\nonumber \\
-\theta^{\mu\sigma}(g^{\nu\rho} r\cdot s-r^\nu s^\rho)
+\theta^{\nu\rho}(g^{\mu\sigma}r\cdot s-r^\sigma s^\mu)+
\theta^{\nu\sigma}(g^{\mu\rho}r\cdot s-r^\mu s^\rho)
\nonumber \\
-\theta^\mu\cdot r \,(s^\nu g^{\rho\sigma}-s^\rho
g^{\nu\sigma})-\theta^\mu\cdot s\,(r^\nu g^{\rho\sigma}-r^\sigma
g^{\nu\rho})+\theta^\nu\cdot r \,(s^\mu g^{\rho\sigma}-s^\rho
g^{\mu\sigma})
\nonumber \\
+\theta^\nu\cdot s \,(r^\mu g^{\rho\sigma}-r^\sigma g^{\mu
\rho})+\theta^\rho\cdot r \,(s^\nu g^{\mu\sigma}-s^\mu
g^{\nu\sigma})+\theta^\rho\cdot s \,(r^\mu g^{\nu\sigma}-r^\nu
g^{\mu\sigma})
\nonumber \\
\left.
+\theta^\sigma\cdot r \,(s^\mu g^{\nu\rho}-s^\nu
g^{\mu\rho})+\theta^\sigma\cdot s \,(r^\nu g^{\mu\rho}-r^\mu
g^{\nu\rho})+r\cdot \theta\cdot s \,(g^{\nu\rho} g^{\mu \sigma}-
g^{\mu \rho} g^{\nu \sigma}) \right\}
\nonumber \\
+\{ (\mu,p,a)\leftrightarrow (\sigma,s,d)\}
+\{ (\rho,r,c) \leftrightarrow (\mu,p,a)\}
+\{ (\sigma,s,d) \leftrightarrow (\nu,q,b) \}
\nonumber \\
+\{ (\rho,r,c) \leftrightarrow (\nu,q,b) \}
+\{(\rho,r,c) \leftrightarrow (\nu,q,b) \mbox{ , } (\sigma,s,d)
\leftrightarrow (\mu,p,a)\} \,\, .
\end{eqnarray}
There are 24 permutations of the four gluon lines in the gggg vertex;
four of these yield the explicit part of the result shown above, while
the rest correspond to the indicated permutations.
Note that the first, third and fourth vertices above provide 
$O(\theta)$ corrections to the corresponding standard model results, 
while the second has no standard model counterpart. These Feynman rules
differ significantly from those quoted in Ref.~\cite{Hinch} for
noncommutative QCD.   For example, only part of the Feynman rule
for the quark-quark-gluon vertex in the construction described
here can be interpreted as a simple phase factor expanded to order
$\theta$. This has phenomenological consequences that we explore in
the next section.

\section{Lorentz Violation}

Noncommutative theories violate Lorentz invariance.  Experimental
searches for Lorentz violation~\cite{Berglund95,searches} allow  one 
to  place limits on the scale of noncommutativity through the effects 
of operators like the one in Eq.~(\ref{eq:theop}).  In this section we 
isolate the most significant Lorentz-violating operators generated 
radiatively in noncommutative QCD, and then study their phenomenological 
implications.

Bounds from Lorentz violation have been studied
previously~\cite{Anis,Moc,Carr,Chaichian:2001si}.  In particular,
operators like the one given above have been considered by Anisimov {\it et
al.}~\cite{Anis} and shown to arise via quantum effects at the two-loop 
level in a theory with noncommutative Yukawa and U(1) gauge interactions.
Mocioiu {\it et al.}~\cite{Moc} have suggested that stronger bounds 
could arise from noncommutative effects in QCD, but did not show how to 
obtain this result from a consistent noncommutative, non-Abelian theory.  
With the formulation of noncommutative QCD suggested by the last section, 
we can explicitly calculate effective operators that violate Lorentz symmetry
and find that they arise already at one loop.

\begin{figure}
\centerline{\epsfbox{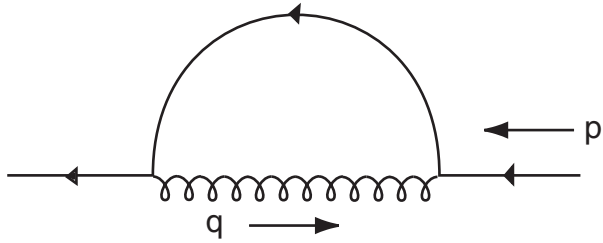}}
\caption{A diagram contributing to 
$\bar q \theta^{\mu\nu} \sigma_{\mu\nu} q$.}
\label{oneloop}
\end{figure}

Figure~\ref{oneloop} shows the one-loop diagram that generates an
effective $\overline{q} \theta^{\mu\nu} \sigma_{\mu\nu} q$ operator 
at lowest order in perturbation theory. (The diagram with a single
$qqgg$ vertex vanishes since this vertex is antisymmetric in its
Lorentz indices, while the gluon propagator is symmetric.)  For 
noncommutative QED with one charge~\cite{Haya}, as well as 
noncommutative U(N)~\cite{Hinch}, the diagram in Fig.~\ref{oneloop} is 
the same as in the commutative case, because the relevant Feynman rules 
give only phases at each vertex, which precisely cancel in the amplitude.   
For noncommutative SU(N) we have

\begin{eqnarray}
i {\cal M} &=& - g^2 T^a T^a \int {(dq) \over q^2 - \lambda^2}  
        \times   
    \bigg\{ \gamma^\mu (1-{i\over 2} p \cdot\theta\cdot q) 
+    {i\over 2}  \theta^{\mu\nu} 
       [(p+q)_\nu (\not\! p -m) -p_\nu (\not\! p + \not\! q -m)] \bigg\}
                                                       \nonumber \\
&\times& {1\over \not\! p + \not\! q -M}  
         \times   
    \bigg\{ \gamma_\mu (1-{i\over 2} q \cdot\theta\cdot p) 
+  {i\over 2}  \theta_{\mu\tau}
     [p^\tau (\not\! p + \not\! q -m) - (p+q)^\tau (\not\! p -m)] \bigg\}
\ .
\end{eqnarray}

\noindent   where $(dq) \equiv d^4 q / (2\pi)^4$.  To evaluate the
diagram as shown, we set $\lambda = 0$ and $M=m$; the  more general
notation is useful since we will employ Pauli-Villars regularization to
handle the divergences. 

Keeping just the $O(\theta)$ terms,

\begin{equation}
i {\cal M}(\lambda^2,M^2) = 
               {2\over 3} g^2 \{ (\not\! p -m), \sigma_{\mu\alpha} \}       
\times \int {(dq)\over (q^2- \lambda^2) 
                         \left( (p+q)^2-M^2 \right)   }
            q^\alpha \theta^{\mu\nu} (p+q)_\nu   \ ,
\end{equation}

\noindent for SU(3). This result is gauge invariant; it would be 
obtained in any generalized Lorentz gauge. The Pauli-Villars 
regulated amplitude is 
${\cal M} \rightarrow {\cal M}(0,m^2) - {\cal M}(\Lambda^2,m^2) - 
{\cal M}(0,\Lambda^2) + {\cal M}(\Lambda^2,\Lambda^2)$, where $\Lambda$
is a large mass scale.  The result is

\begin{equation}                             \label{result}
{\cal M} = {g^2 \over 96 \pi^2} \bigg(
       \big\{ (m - \not\! p),  \ 
       \Lambda^2 \theta^{\mu\nu} \sigma_{\mu\nu}   \big\} 
-{2\over 3} \big\{ (m - \not\! p),  \ 
       p_\mu \theta^{\mu\nu} \sigma_{\nu\tau} p^\tau \ln{\Lambda^2}  
                         \big\}  \bigg) \ , 
\end{equation}

\noindent for the term leading in $\Lambda$ for each Dirac structure.  The
same result follows if one just cuts off the integral in
${\cal M}(0,m^2)$ at a large (Euclidean) $q^2 = \Lambda^2$, the approach
that was adopted in~\cite{Anis}.

We conclude that at lowest order in perturbation theory, the formulation
of noncommutative QCD that we have described leads to the set of Lorentz
violating operators

\begin{equation}
m \theta^{\mu\nu} \bar q \sigma_{\mu\nu} q , \,\,\,\,\,  
\theta^{\mu\nu} \bar q \sigma_{\mu\nu} \not\!\! D q , \,\,\,\,\,
\mbox{ and } \,\,\,\,\,
\theta^{\mu\nu} D_\mu \bar q \ \sigma_{\nu\rho} D^\rho q,  
\end{equation}

\noindent with coefficients as given by Eq.~(\ref{result}). Note
that an operator with three derivatives can be eliminated using
$\{\not\! p \mbox{ , }\sigma_{\nu\tau} p^\tau\}=0$.

Phenomenological bounds follow because the noncommutativity matrix
$\theta^{\mu\nu}$ has a fixed orientation. The most significant
phenomenological bound comes from the first term in Eq.~(\ref{result}),
which corresponds to the leading divergence.  Part of
$\theta^{\mu\nu} \sigma_{\mu\nu}$ acts like a 
$\vec\sigma \cdot \vec B$ interaction with a fixed $\vec B$, which
leads to sidereal variations in, for example, hyperfine energy
splittings.  Such variations in the differences between Cs and Hg
atomic clocks, which have different sensitivities to an external 
$\vec\sigma \cdot \vec B$-like interaction, are bounded at the
$10^{-7}$ Hz, or few
$\times$
$10^{-31}$ GeV, level~\cite{Berglund95}.  

We estimate our operator in a nucleon or nuclear environment.  The
up and down quarks are very light, but are off shell by about the mass of
a ``constituent quark,'' and we take
$
p^0 \approx m_{\rm cons} \approx 300 {\rm\ MeV}$ and
$
\vec p \approx 0$.  Then,

\begin{equation}\label{eq:3.7}
{\alpha_s \over 12\pi} \Lambda^2 \theta \, m_{\rm cons} < \Delta E  \ , 
\end{equation}

\noindent where $\theta$ is a typical scale for elements of the
matrix $\theta^{\mu\nu}$ and $\Delta E$ is a bound on the sidereal
variation of an energy difference.  For the atomic clock bound cited
above, and using $\alpha_s =1$, one finds

\begin{equation}
\theta \Lambda^2 \mbox{\raisebox{-1.0ex} 
{$\stackrel{\textstyle ~<~} {\textstyle \sim}$}} 10^{-29}  \ .
\end{equation}

\section{Discussion}

We have derived the Feynman rules for a consistent formulation of 
noncommutative QCD and have used them to compute the most dangerous, 
Lorentz-violating operator that is generated through radiative corrections. 
While this operator vanishes when quarks are on shell, those in a nucleon 
are typically off shell by an amount comparable to a constituent quark 
mass, and thus the matrix element between 
nucleon states should remain nonvanishing.  From here we obtained the 
approximate bound 
\begin{equation}\label{eq:bound}
\theta \Lambda^2 \mbox{\raisebox{-1.0ex} 
{$\stackrel{\textstyle ~<~} {\textstyle \sim}$}} 10^{-29} \,\,\, ,
\end{equation}
where $\Lambda$ is an ultraviolet regularization scale. This result follows 
from tests of Lorentz invariance in clock-comparison experiments, sensitive 
to the sidereal variation in the relative Larmor precession frequencies of 
Hg and Cs~\cite{Berglund95}. The calculational uncertainty in our analysis 
lies in the extrapolation of our perturbative result to the nonperturbative 
regime in which nucleon magnetic moments and hadronic matrix elements are 
evaluated.  While we do not address this issue directly here, it is clearly 
reasonable to assume that our nonvanishing result changes smoothly as the 
QCD coupling is increased, and that the estimate of Eq.~(\ref{eq:3.7})
gives a reasonable indication of the magnitude of the effect. This has been 
the approach adopted almost uniformly in the literature~\cite{Anis,Moc}, 
pending a more detailed strong interaction calculation. We have also shown 
that our result persists in two different regularization schemes (and no 
doubt will appear in others).  Even allowing for uncertainty in how the 
regularization scale is best defined, it is clear that if 
$\Lambda \sim 1$~TeV (as one would expect if the Planck scale is low) 
then Eq.~(\ref{eq:bound}) requires $\Lambda_{{\rm NC}}\mbox{\raisebox{-1.0ex} 
{$\stackrel{\textstyle ~>~} {\textstyle \sim}$}} 10^{17}$~GeV, which is far 
above the cutoff.  This observation, or equivalently the appearance of an 
extremely small dimensionless number in the theory, Eq.~(\ref{eq:bound}), is 
an indication of the unnaturalness of spacetime noncommutativity, as it is 
presently defined.

{\samepage
\begin{center}
{\bf Acknowledgments}
\end{center}
C.E.C. and C.D.C. thank the National Science Foundation for support
under Grant No.\ PHY-9900657. In addition, C.D.C. thanks the
Jeffress Memorial Trust for support under Grant No.~J-532. R.F.L.
thanks the Department of Energy for support under Grant 
No.\ DE-AC05-84ER40150.  We thank P. Amore, A. Aranda, A. Armoni, 
A. Kosteleck\'{y}, and N. Zobin for useful communications.}


\end{document}